\documentclass[twocolumn,showpacs,preprintnumbers,amsmath,amssymb]{revtex4}

\usepackage{graphicx}
\usepackage{dcolumn}
\usepackage{bm}
\usepackage{dsfont}
\begin{document}  
\title{Excited hadrons on the lattice: Baryons}
\author{Tommy Burch$^a$, 
Christof Gattringer$^b$, 
Leonid Ya.\ Glozman$^b$, 
Christian Hagen$^a$, 
Dieter Hierl$^a$,
C.\ B.\ Lang$^b$, and 
Andreas Sch\"afer$^a$\\
(BGR [Bern-Graz-Regensburg] Collaboration)}
\vskip1mm
\affiliation{$^a$ Institut f\"ur Theoretische Physik, Universit\"at
Regensburg, D-93040 Regensburg, Germany}
\affiliation{$^b$ Institut f\"ur Physik, FB Theoretische Physik, Universit\"at
Graz, A-8010 Graz, Austria}
\begin{abstract}
We present results for masses of excited baryons from a quenched calculation 
with Chirally Improved quarks at pion masses down to 350 MeV. 
Our analysis of the correlators is based on the variational method. 
In order to provide a large basis set for spanning the physical states, we use
interpolators with different Dirac structures and Jacobi smeared quark
sources of different width. Our spectroscopy results for a wide range of
ground state and excited baryons are discussed.

\end{abstract}
\maketitle

\section{Introduction}

The reproduction of the hadron mass spectrum from first principles 
is an important challenge for lattice QCD. Ground state spectroscopy 
on the lattice
is by now a well understood problem and impressive agreement with
experiments has been achieved. However, the lattice study of excited
states \cite{excitedspectro}--\cite{exmeson} 
is not as advanced. The reason for this is twofold: Firstly, the
masses of excited states have to be extracted from subleading exponentials in
the spectral decomposition of two-point functions. Secondly, 
the construction of
hadron interpolators which have a good overlap with the wave functions
of excited states is much more challenging than for the ground state.

Concerning the first issue, the extraction of the signal from the subleading
exponential, several approaches such as constrained fitting or the maximum
entropy method can be found in the literature
\cite{constrainedfits,maximumentropy}. 
Here we apply the variational method \cite{Mi85,LuWo90}, 
where not only a single correlator is analyzed, but a
matrix of correlators is used. This matrix is built from several different
interpolators, all with the quantum numbers of the desired state. 
The variational method also incorporates in a natural way 
a solution to the second issue, the wave function of the excited states: 
One uses a set of basis
interpolators which is large enough to span ground and excited states and the
variational method finds the optimal combinations of them. 
In principle, no prior knowledge of or
assumption about the composition of the physical hadron state has to be used. 

However, the variational method can succeed only if the basis set of hadron
interpolators is rich enough to span ground and excited states. On the other
hand, the basis should also be constructed such that it can be
implemented numerically in an efficient way without the need for many
different quark sources. In this article we use a twofold strategy for
building our basis interpolators: We use interpolators with different Dirac 
structures and furthermore compose them using different types of 
smearing for the
individual quarks. In particular, we apply different amounts of Jacobi
smearing \cite{jacobi} and in this way create ``narrow'' and ``wide'' 
sources. A
combination of these allows for spatial wave functions with nodes, which are
essential for a good overlap with excited states.  

Following a first test of the outlined strategy \cite{firsttest} 
and an analysis of mesons
with our method \cite{exmeson}, in this paper we present 
in detail the results obtained for
baryons. In the next section we collect the basic equations 
for the implementation of
the variational method, detail the construction of our sources and give an
overview of the parameters of our numerical simulation. Subsequently we
discuss effective mass plots, the eigenmodes of the correlation matrix, as
well as the baryon masses and their chiral extrapolations. The paper 
closes with a summary and an outlook. 

\section{Outline of the calculation}

\subsection{The variational method}

As already stated, we use the variational method \cite{Mi85,LuWo90} 
to extract the masses of ground
and excited states. Starting from a set of basis operators $O_i, \, i = 1,2\,
... \, N$, we compute the correlation matrix
\begin{equation}
C_{ij}(t) \; = \; \langle \, O_i(t) \, \overline{O}_j(0) \, \rangle \; . 
\label{corrmatdef}
\end{equation}
In Hilbert space these correlators have the decomposition 
\begin{equation}
C_{ij}(t) \; = \; \sum_n \langle \, 0 \, | \, O_i \, | \, n \, \rangle 
\langle \, n \, | \, O_j^\dagger \, | \, 0 \, \rangle \, e^{-t \, M_n}  \; . 
\label{corrmatrix}
\end{equation}
Using the factorization of the amplitudes one can show \cite{LuWo90} 
that the eigenvalues 
$\lambda^{(k)}(t)$ of the generalized eigenvalue problem 
\begin{equation}
C(t) \, \vec{v}^{\,(k)} \; \; = \; \; \lambda^{(k)}(t) \, C(t_0) \, 
\vec{v}^{\,(k)} \; , 
\label{generalized}
\end{equation}
behave as 
\begin{equation}
\lambda^{(k)}(t) \; = \; e^{-(t-t_0) \, M_k} \,[ \, 1 +
{\cal O}(e^{-(t-t_0) \, \Delta M_k}) \,] \; , 
\label{eigenvaluedecay}
\end{equation}
where $M_k$ is the mass of the $k$-th state and $\Delta M_k$ is the difference 
to the masses of neighboring states. In Eq.\ (\ref{generalized}) 
the eigenvalue problem is normalized with respect to a timeslice $t_0 < t$.

At this point we remark, that the variational method can be generalized to
include also ghost contributions as they appear in a quenched or partially
quenched calculation. The fact that ghost contributions play a role also 
for baryons was first stressed in \cite{excitedspectro}. 
In the spectral decomposition (\ref{corrmatrix}) ghosts
appear with a modified time dependence, possibly including also a negative
sign. In \cite{ghostvar} it was shown that the ghost contribution couples 
to an individual 
eigenvalue (up to the correction term) in the same way as a ``proper''
physical state. Thus, ghost contributions are disentangled from the desired states
and need not be modeled in the further analysis of the
exponential decay of the eigenvalues.

Let us finally stress that also the eigenvectors of the generalized eigenvalue
problem (\ref{generalized}) contain interesting information. 
If one plots the entries of the
eigenvector as a function of $t$, one finds that they are essentially constant
in the same range of $t$-values where plateaus of the effective mass are seen
(compare Fig.\ \ref{evec_vs_t}). These plateaus can be used to optimize the
interval for fitting the eigenvalues. Furthermore, the eigenvectors encode 
the information which
linear combinations of the basis interpolators couple to which eigenvalue and
thus provide one with a ``fingerprint'' of the corresponding states. Comparing
these fingerprints for different values of the quark mass is an important
cross-check for the correct identification of the states.

\subsection{Dirac structure of the baryon interpolators}   

For the baryons we analyze, we use the following
operators with different Dirac structures:
\begin{itemize}
\item
Nucleon:
\begin{eqnarray}
N^{(i)} & = & \epsilon_{abc} \Gamma^{(i)}_1 u_a ( u_b^T  \Gamma^{(i)}_2 d_c -
d_b^T  \Gamma^{(i)}_2 u_c ) \; .
\label{nucleoninterpolator}
\end{eqnarray}
\item
$\Sigma$:
\begin{eqnarray}
\Sigma^{(i)} & = & \epsilon_{abc} \Gamma^{(i)}_1 
u_a ( u_b^T  \Gamma^{(i)}_2 s_c - s_b^T  \Gamma^{(i)}_2 u_c ) \; .
\end{eqnarray}
\item
$\Xi$:
\begin{eqnarray}
\Xi^{(i)} & = & \epsilon_{abc} \Gamma^{(i)}_1 
s_a ( s_b^T  \Gamma^{(i)}_2 u_c - u_b^T  \Gamma^{(i)}_2 s_c ) \; .
\end{eqnarray}
\item
$\Lambda$-octet:
\begin{eqnarray}
\Lambda_8^{(i)} & = & \epsilon_{abc} \{ \Gamma^{(i)}_1 
s_a ( u_b^T  \Gamma^{(i)}_2 d_c - d_b^T  \Gamma^{(i)}_2 u_c ) \\ \nonumber
& & \,\, + \; \Gamma^{(i)}_1 u_a ( s_b^T  \Gamma^{(i)}_2 d_c) - 
\Gamma^{(i)}_1 d_a ( s_b^T  \Gamma^{(i)}_2 u_c) \; .
\label{octetinterpolator}
\end{eqnarray}
\item
$\Lambda$-singlet:
\begin{eqnarray}
\Lambda_1 & = & \epsilon_{abc} \Gamma^{(1)}_1 u_a ( d_b^T \Gamma^{(1)}_2 s_c - 
s_b^T \Gamma^{(1)}_2 d_c) \\ \nonumber
&  & \,\,+ \; \mbox{cyclic permutations of}\;  u, d, s  \; .
\end{eqnarray}
\item
$\Delta$:
\begin{eqnarray}
\Delta_\mu & = & \epsilon_{abc}  u_a ( u_b^T C\gamma_\mu  u_c ) \; .
\label{deltainterpolator}
\end{eqnarray}
\item
$\Omega$:
\begin{eqnarray}
\Omega_\mu & = & \epsilon_{abc}  s_a ( s_b^T C\gamma_\mu  s_c ) \; .
\label{omegainterpolator}
\end{eqnarray}
\end{itemize}
Here we used vector/matrix notation for the Dirac indices. The different
possible choices 
for $\Gamma^{(i)}_1$ and $\Gamma^{(i)}_2$ are listed in Table \ref{gammatable}.

\begin{table}[t]
\begin{center}
\begin{tabular}{c|c|c}
\hspace{10mm} & \hspace{7mm}$\Gamma^{(i)}_1$\hspace{7mm} & 
\hspace{7mm}$\Gamma^{(i)}_2$\hspace{7mm} \\
\hline
$i=1$ & $\mathds{1}$  & $C\gamma_5$         \\
$i=2$ & $\gamma_5$    & $C$                 \\
$i=3$ & $i\mathds{1}$ & $C\gamma_4\gamma_5$ \\
\end{tabular}
\end{center}
\caption{Dirac structures used for nucleon, $\Sigma,\, \Xi$ and
 $\Lambda$-octet, according to Eqs.\ 
(\ref{nucleoninterpolator})--(\ref{octetinterpolator}).}
\label{gammatable}
\end{table}

Our interpolator for the $\Delta$ ($\Omega$) still has overlap with both spin-$\frac12$
and spin-$\frac32$. Thus, we need a projection to definite 
angular momentum. We use the continuum formulation of a 
spin-$\frac32$ projection for a Rarita-Schwinger field:
\begin{eqnarray*}
P^{3/2}_{\mu \nu} (\vec{p}) & = & \delta_{\mu \nu} - \frac{1}{3} 
\gamma_{\mu} \gamma_{\nu} - \frac{1}{3p^2} ( \gamma \cdot p \, 
\gamma_{\mu} p_{\nu} + p_{\mu}\gamma_{\nu} \gamma \cdot p),
\end{eqnarray*}
where $p_\mu$ is the 4-momentum, in our case given by $(\vec{0},m)$.
For each component of the projected $\Delta$ ($\Omega$) we compute the
correlator and average these 2-point functions over $\mu,\nu = 1,2,3$.

Finally, our baryon correlators are 
projected to definite parity using the projection operator 
$P^\pm=\frac12(\mathds{1} \pm \gamma_4)$. We obtain two matrices 
of correlators: 
\begin{eqnarray}
C^+_{ij}(t) & = &  Z_{ij}^+ e^{-tE^+} + Z_{ij}^- e^{-(T-t)E^-} ,
\end{eqnarray}
where we have projected with $P^+$ and 
\begin{eqnarray}
C^-_{ij}(t) & = &  - Z_{ij}^- e^{-tE^-} - Z_{ij}^+ e^{-(T-t)E^+} ,
\end{eqnarray}
when using $P^-$.
These two matrices are combined to
\begin{eqnarray}
C(t) & = & \frac{1}{2} \left( C^+(t) - C^-(T-t) \right) \; ,
\end{eqnarray}
to improve statistics. This gives rise to the final correlator $C(t)$
which we then use in the variational method. The positive
parity states are obtained from the correlator at small $t$ running forward
in time, while the negative parity states are found at large time 
arguments, propagating backward in
time with $T-t$. As expected, the correlation matrices $C(t)$ are real 
and symmetric within error bars and
we therefore symmetrize the matrices by replacing $C_{ij}(t)$ with
$C_{ij}(t) = [C_{ij}(t) + C_{ji}(t)]/2$ before diagonalization. 

\subsection{Quark sources}

In addition to the different Dirac structures, we construct the interpolators
listed in the last section from quarks with sources created by different
amount of smearing. In particular, 
we use Jacobi smearing \cite{jacobi} with two different sets of parameters 
(number of
smearing steps, amplitude of hopping term) to create narrow and wide sources. 
The shapes of these sources are approximately gaussian with $\sigma \sim 0.27$
fm for the narrow source and $\sigma \sim 0.41$ fm for the wide
source. Details of the source preparation and plots of the source shapes can
be found in \cite{firsttest,exmeson}.

Each quark in our baryon interpolators can either be narrow ($n$) or wide
($w$), giving rise to the following eight combinations for the sources:
\begin{eqnarray}
&& n(nn)\,,\, n(nw)\,,\, n(wn)\,,\, w(nn)\,, 
\nonumber \\
&& n(ww)\,,\, w(nw)\,,\, w(wn)\,,\, w(ww) \; .
\end{eqnarray}
In this notation the order of the quark fields is understood as in Eqs.\
(\ref{nucleoninterpolator}) -- (\ref{omegainterpolator}) and the parentheses
indicate which quarks are combined in the diquark combination. 
Since the smearing used here is a purely scalar operation, the 
assignment of quantum numbers, as given
in the last subsection, remains unchanged. 

Taking into account the
different Dirac structures discussed in the last section, we can work with 24
different interpolators for nucleon, $\Sigma$, $\Xi$ and $\Lambda$-octet. For 
the $\Lambda$-singlet and the $\Delta$ we have only one Dirac structure and
consequently a total of 8 different interpolators. 
We remark that in the final analysis not all interpolators are used. We prune
the maximal correlation matrix and remove some of the correlators that couple
only weakly to the physical states or add no new information,
thus enhancing the numerical noise. The criterion for the
selection of the interpolators is the optimization of the 
quality of the plateaus in the effective mass
\begin{equation}
aM_{eff}^{(k)} \, \Big(t + \frac{1}{2}\Big) \; = \; 
\ln \left( \frac{\lambda^{(k)}(t)}
{\lambda^{(k)}(t+1)} \right) \; .
\label{effmass}
\end{equation}

\subsection{Parameters of the simulation}

We work with quenched gauge configurations generated with the L\"uscher-Weisz
action \cite{luweact}. We use two sets of lattices, 
$20^3\times 32$ and $16^3 \times 32$, at
couplings $\beta = 8.15$ and $\beta = 7.90$ corresponding to
lattice spacings of $a =
0.119$ fm and $a = 0.148$ fm, determined from the Sommer parameter in
\cite{scale}. Thus for both lattices we have a spatial extent of $L \sim 2.4$
fm. The
two different values of the lattice constant $a$ allow us to assess the
cutoff dependence. The parameters of the gauge configurations are collected in
Table \ref{simparameters}.
  
Our quark propagators are computed using the Chirally Improved (CI) Dirac
operator \cite{cioperator}. The CI operator is a systematic approximation 
of a solution of the
Ginsparg-Wilson equation \cite{giwi} with good chiral properties
\cite{citests}. 
We work with
several quark masses in the range $am = 0.02,\, ... \, 0.2$, 
leading to pion
masses down to 350 MeV. For setting the strange quark mass we use the K-meson
with the light quark mass extrapolated to the chiral limit. 

Our quark sources are placed at $t = 0$ and the generalized eigenvalue problem
(\ref{generalized})  is normalized at $t_0 = 1a$. The final results for the
baryon masses were obtained from a fully correlated fit to the eigenvalues. 
The errors we show are statistical errors determined with single elimination
jackknife.

\begin{table}[t]
\begin{center}
\begin{tabular}{c|c|c|c|c}
\hspace{3mm} size \hspace{3mm} & \hspace{3mm} $\beta$ \hspace{3mm} & 
\hspace{3mm} confs.\hspace{3mm} & \hspace{3mm} $a$[fm] \hspace{3mm} &
\hspace{3mm} $a^{\!-1}$[MeV] \hspace{3mm}\\
\hline 
$20^3\! \times \! 32$ & 8.15 & 100 & 0.119 & 1680  \\
$16^3\! \times \! 32$ & 7.90 & 100 & 0.148 & 1350  \\
\end{tabular}
\end{center}
\caption{
Parameters of our simulation. We list the lattice size, the inverse coupling 
$\beta$ the number of configurations, the lattice spacing $a$ and the 
cutoff $a^{-1}$.}
\label{simparameters}
\end{table}

\section{Results}

\subsection{Effective masses, eigenvectors and fit ranges}

\begin{figure*}[t]
\begin{center}
\includegraphics[width=\textwidth,clip]{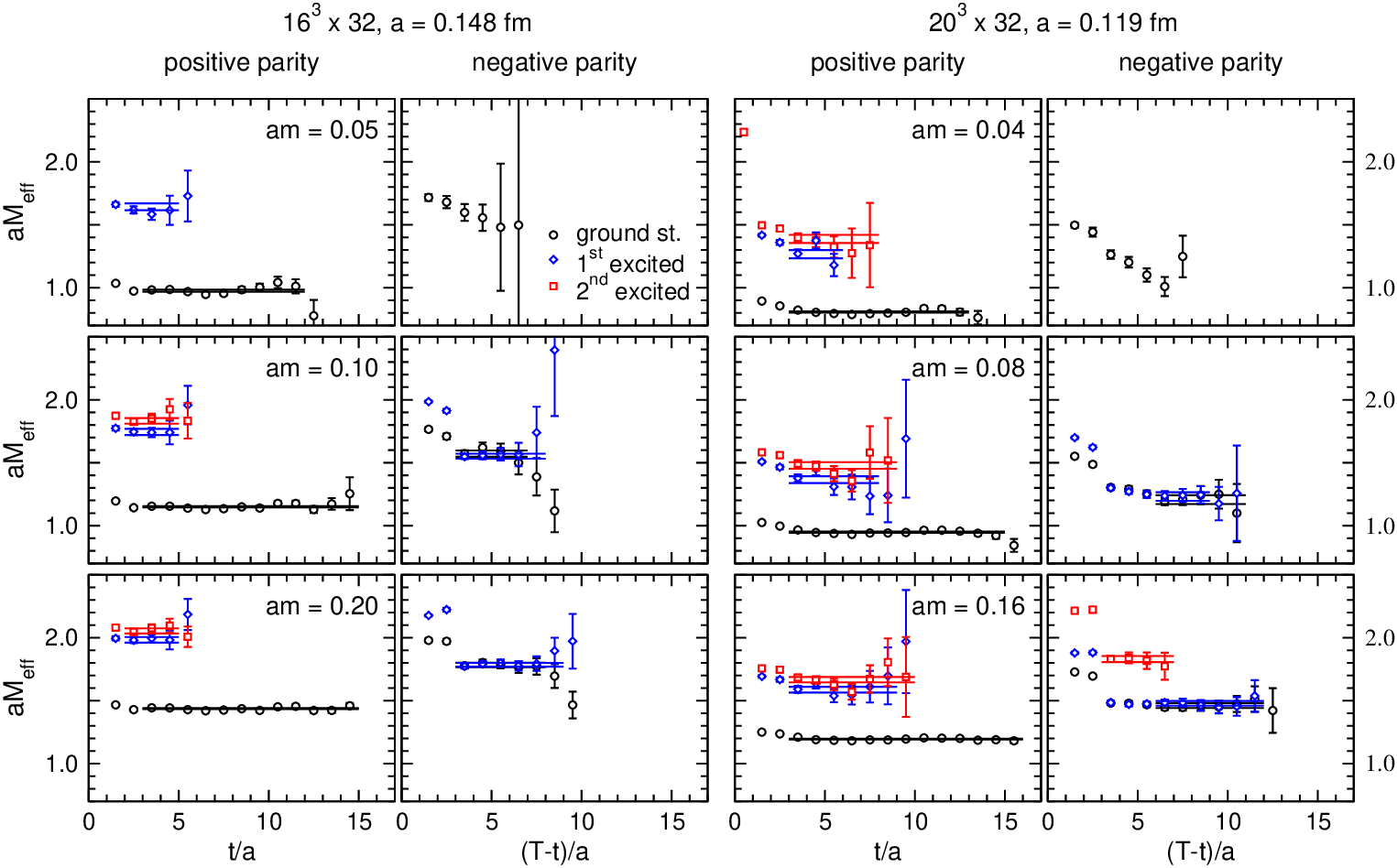}
\end{center}
\caption{Effective mass plots for nucleon ground and excited states. 
We compare the results from our coarse ($16^3 \times 32$, $a=0.148$
fm, $am_q=0.05,0.1,0.2$ top to bottom), and fine ($20^3 \times 32$, 
$a=0.119$ fm, $am_q=0.04,0.08,0.16$) lattices.  The solid lines are
the results from correlated fits of the eigenvalues. They represent 
the fit result plus and minus the corresponding error.
}
\label{effmasspos}
\end{figure*}

\begin{figure}[t]
\vspace*{-6mm}
\hspace*{-1mm}
\includegraphics[width=8.0cm,clip]{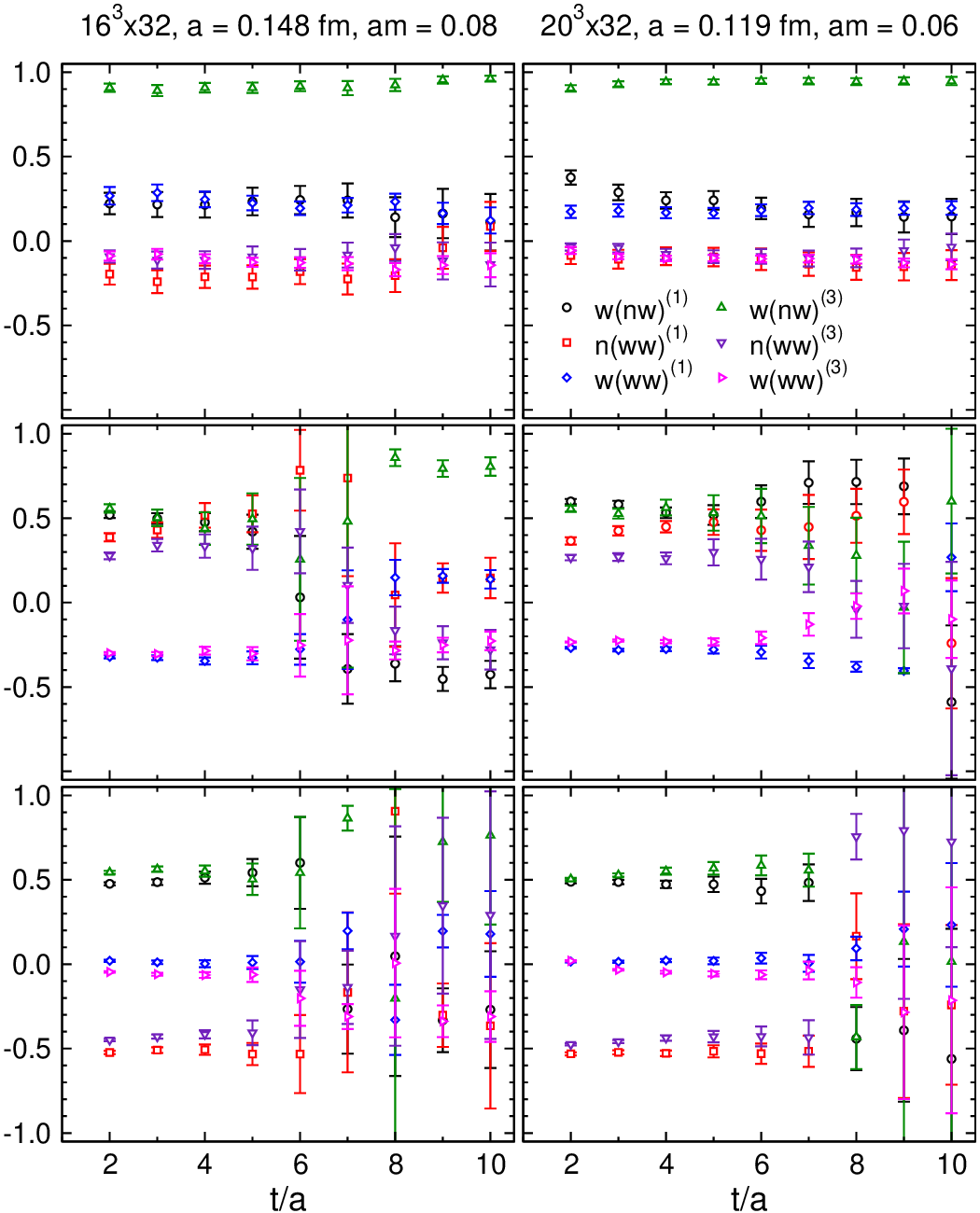}
\caption{Eigenvectors for nucleon ground and excited positive parity
states. From top to bottom we show the eigenvector components of ground,
first and second excited  state.}
\label{evec_vs_t}
\end{figure}
Let us begin our presentation with a discussion of effective masses 
(\ref{effmass}) for
the nucleon system. For positive parity the combination of the 6 operators
$n(ww)^{(1)}$, $w(wn)^{(1)}$, $w(ww)^{(1)}$, $n(ww)^{(3)}$, $w(wn)^{(3)}$, 
$w(ww)^{(3)}$, (the upper index denotes the choice of Dirac structures
according to Table \ref{gammatable}) gives the strongest signal. For negative
parity we used the $4 \times 4$ correlation matrix built from 
$n(nn)^{(1)}$, $w(nn)^{(1)}$, $n(nn)^{(2)}$, $w(nn)^{(2)}$. 

In Fig.\ \ref{effmasspos} we compare the effective mass plots for positive and
negative parity nucleons from our two 
lattices at different values of the bare quark
mass; $am = 0.05,0.1,0.2$ for $16^3\times 32$ and $am = 0.04, 0.08, 0.16$ for
$20^3 \times 32$. These numbers were chosen such that they give rise to
approximately equal pion masses for the two lattice spacings used. The plots
also contain the nucleon masses in lattice units as obtained from a correlated 
fit of the propagator (horizontal bars giving the central value plus and 
minus the statistical error).
The figure shows clear long plateaus for the ground state masses, while the 
signals for excited states have larger error bars and shorter plateaus. 
Furthermore the quality of the data decreases as the quarks become
lighter -- a feature well known in lattice spectroscopy.  

Another important piece of information comes from the eigenvectors. In Fig.\ 
\ref{evec_vs_t} we show the 6 entries of the lowest three 
eigenvectors corresponding to ground, first and second excited state
(top to bottom) in the positive parity nucleon channel. 
Again we compare the results for our two lattice sizes using quark 
mass values which give rise to essentially the same pion mass. 
For each value of $t$ the respective eigenvectors are normalized to unit length. 

It is interesting to note that the eigenvectors are only weakly dependent on 
$t$ (actually this can be shown from the generalized eigenvalue 
problem). The entries form plateaus which are very long for the ground states 
but also for the excited states often contain 4 to 8 values of $t$.
Typically these plateaus extend at least over the same number of 
$t$-values as the effective mass plateaus -- often they are even longer by 
one or two points. 

As in the case of effective masses, 
the formation of the eigenvector 
plateaus indicates that the channel is dominated by a
single state. Thus, the eigenvector plateaus provide an important tool for the
reliable identification of the $t$-intervals where the eigenvalues
can be used for a fit. 
Indeed, sometimes it is the eigenvectors which prevent one from fitting  
``quasiplateaus'' in the effective mass.
Due to relatively large statistical errors in the effective masses,
the data sometimes resemble a plateau and it is only the
absence of a plateau in the corresponding eigenvectors
which allows us to conclude that a quasiplateau is not conclusive.
We implemented this strategy and now fit the eigenvalues
only where we see also eigenvector plateaus. 

We finally remark that the values for the eigenvectors are almost
exactly the same for the two values of the cutoff we consider (the left hand
side plots are for $a = 0.148$ fm, the right hand side is for $a = 0.119$). 
Although the entries of the eigenvectors cannot be expected to scale (they are
linear combinations of matrix elements of our interpolators with the physical
states), it is reassuring for the application of the method that no large 
discrepancies are observed.

\subsection{Nucleon}

\begin{figure*}[t]
\begin{center}
\includegraphics*[height=70mm,clip]{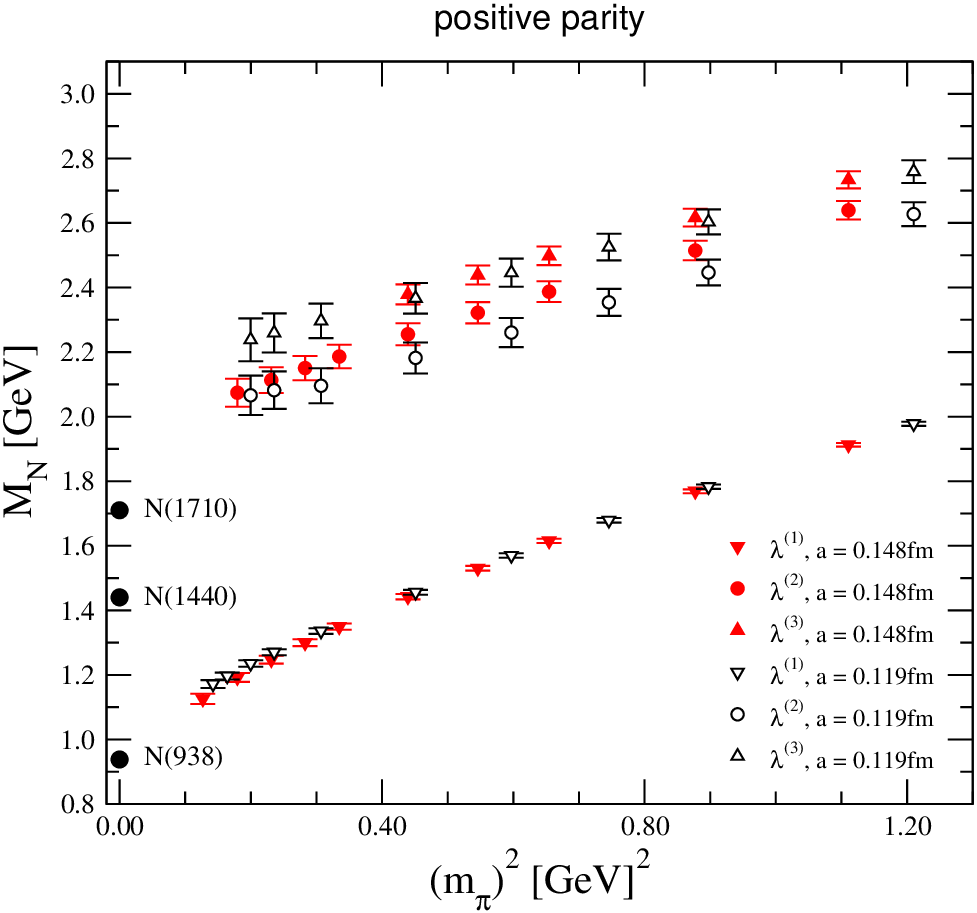}
\hspace*{10mm}
\includegraphics*[height=70mm,clip]{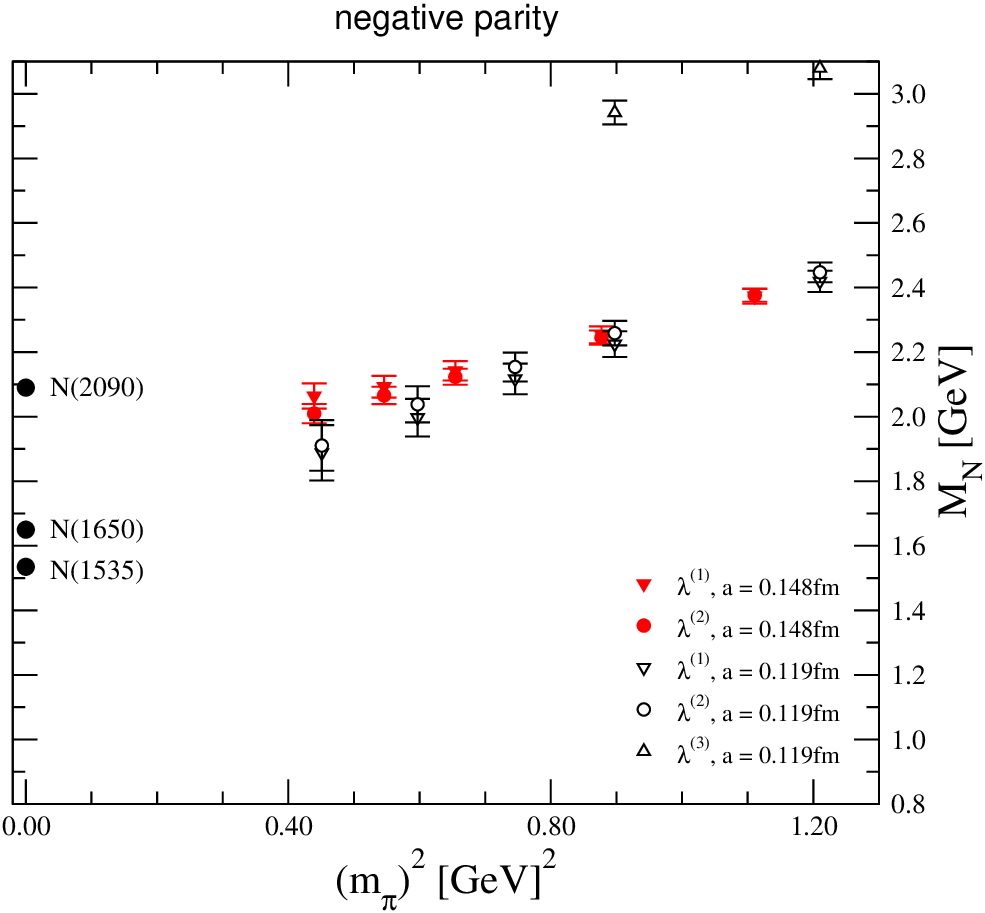}
\end{center}
\caption{
Ground and excited state nucleon masses versus $M_\pi^2$ for 
our two lattices. Filled symbols are used for $16^3\times32, \, a = 0.148$ fm,
open symbols for $20^3\times32, \, a = 0.119$ fm. The left hand side plot 
shows the positive parity states, the right hand side is for negative parity.
The experimental data are included as filled circles.}
\label{nucleonresults}
\end{figure*}

As already discussed in the last section, the positive parity nucleon 
masses were extracted from the $6 \times 6$ correlation matrix of
$n(ww)^{(1)}$, $w(wn)^{(1)}$, $w(ww)^{(1)}$, $n(ww)^{(3)}$, $w(wn)^{(3)}$, 
$w(ww)^{(3)}$, while for negative parity the $4 \times 4$ correlation 
matrix built from $n(nn)^{(1)}$, $w(nn)^{(1)}$, $n(nn)^{(2)}$, $w(nn)^{(2)}$
was used. Of course these combinations were used for all quark masses.
For the positive parity ground state we could determine the mass 
for all our quark masses. For the excited nucleon states of positive parity 
the combined assessment of effective masses
and eigenvector plateaus did not allow for a trustworthy extraction of the
corresponding nucleon masses for the two smallest quark masses.

We identify two excited states of positive parity which
have not very different masses for the whole 
quark mass region where we see a signal. This is consistent with 
our previous observation on a smaller lattice \cite{firsttest}. 
These are two physically distinct states since they are 
observed in different
eigenvalues of the correlation matrix and the corresponding
eigenvectors are orthogonal. Some additional efforts are
required to properly identify the nature of our quenched
excited states. We follow the strategy of Ref.\ \cite{firsttest},
i.e., we trace the states from the heavy quark region towards the 
physical limit.

In the heavy quark region, where we obtain the best signals,
the quenching and chiral symmetry effects are less important and the
naive quark picture is adequate. Then we know a-priori, that
there must be two approximately degenerate excited states of
positive parity. The first one is a member of the 56-plet of SU(6)
and the second one belongs to the 70-plet. In the excited
56-plet state, as well as in the ground state 56-plet (the nucleon),
all possible quark pairs have positive parity. Then it follows
that the signal from the nucleon, as well as from the excited
56-plet state, can be seen with those interpolators that contain
two-quark subsystems of positive parity (these are the ones
with $i=1,3$ from the Table I). On the other hand,
the positive parity 70-plet state 
contains both positive and negative parity two-quark subsystems,
and can be seen with the $i=2$ interpolator, where the ``diquark'' 
has negative parity. This picture is confirmed in the heavy quark 
limit of our results. If we construct our correlation matrix with
the $i=1$ and/or $i=3$ interpolators, we find both the ground
state (the nucleon) and two excited states of positive parity,
while only one state is observed with the $i=2$ interpolator. This
state corresponds to the positive parity excited state.

Using the ``fingerprint'' from the eigenvectors, we are able to 
trace these states from the heavy quark region,
where their physical nature can be safely identified, to the light
quark region (down to $m_\pi = 450$ MeV), where they still remain
approximately degenerate. Clearly these signals, extrapolated
to the physical region, remain essentially higher than the
experimental states N(1440) and N(1710)
(cf.\ Fig.~\ref{nucleonresults}). Phenomenologically
the latter states are ascribed to the 56-plet and 70-plet,
respectively. 

The discrepancy between our results and the experimental numbers
is probably partly due to quenching, where a significant part of chiral
physics is absent. Also finite volume effects cannot be excluded
(our physical volume is 2.4 fm and large finite volume effects can 
be anticipated for excited states \cite{excitedspectro2}).

Note that the perturbative gluon exchange between valence quarks,
characteristic of the naive constituent quark model,
is adequately represented in the quenched calculation. The
discrepancy of our results with the experimental ones hints that
it is the chiral physics,  partly missing in quenched QCD,
that could shift both positive
parity excited states (and especially the Roper state) down \cite{GR}.

Our results for the nucleons are presented in Fig.\ \ref{nucleonresults}.
The left plot is for positive parity, the right for negative
parity. Filled symbols are used for the $16^3 \times 32, \, a = 0.148$ fm
lattice, open symbols for $20^3 \times 32, \, a = 0.119$ fm. The filled
circles represent the experimental masses. 

The results for the positive parity ground state (left plot,
downward pointing triangles)
agree well with the experimental value (for
the chiral extrapolation of our data see Subsection F). Furthermore, the data
show almost no cutoff effects. For the first excited state
(circles) the results for the two values of the cutoff differ by
about one sigma, while for
the second excited state (upward pointing triangles) the two data sets agree. 
However, both excited states extrapolate to values about 20-30\% larger 
than the experimental numbers. 

\begin{figure}[b!]
\hspace*{-1mm}
\includegraphics[width=8.5cm,clip]{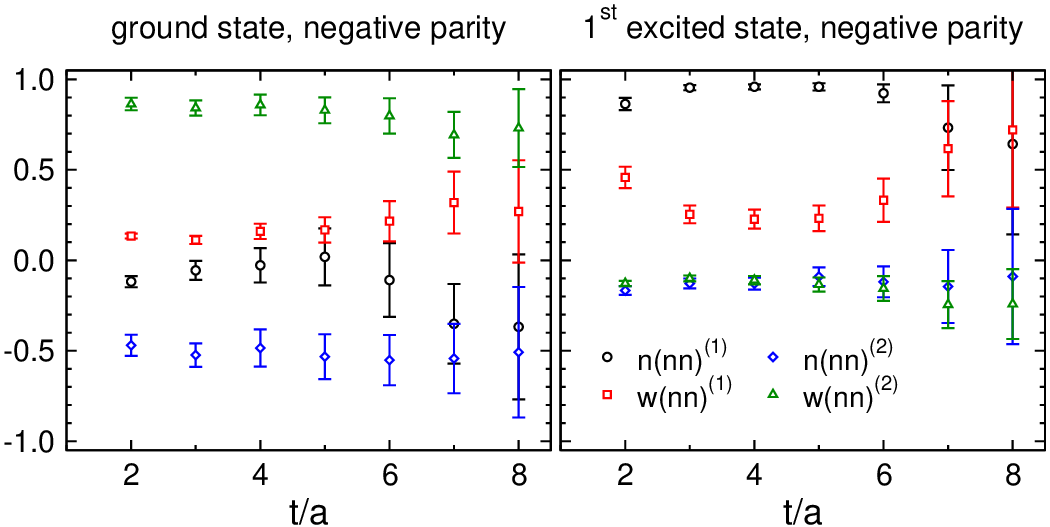}
\caption{Eigenvectors for nucleon ground and first excited negative parity
states. The data are for our $20^3\times 32$ lattice at $am = 0.06$.}
\label{evec_vs_t_nuc_neg}
\end{figure}

For negative parity we mainly fit ground and first excited states. 
Only for the two largest quark masses on the finer lattice we can extract the
second excited mass. We find that the lowest two states are nearly degenerate,
but extrapolate to the physical masses within error bars 
(compare Subsection F). Cutoff effects are clearly 
seen only for small quark masses. 
Since the negative parity ground and first excited state are nearly
degenerate, we checked that they are indeed different by inspecting the
eigenvectors and following their behavior down from the heavy quark region.
Entries of the eigenvectors at quark mass $am = 0.06$ are shown for our 
$20^3\times 32$ lattice  in Fig.\ \ref{evec_vs_t_nuc_neg}. 
In contrast to the positive
parity excited states, the negative parity states fit the experimental data well.
This is expected since the negative parity states have the
mixed flavor-spin symmetry $[21]_{FS}$ and experience only small 
chiral effects \cite{GR}.

\subsection{$\Sigma$ and $\Xi$}

\begin{figure*}
\begin{center}
\includegraphics*[height=70mm,clip]{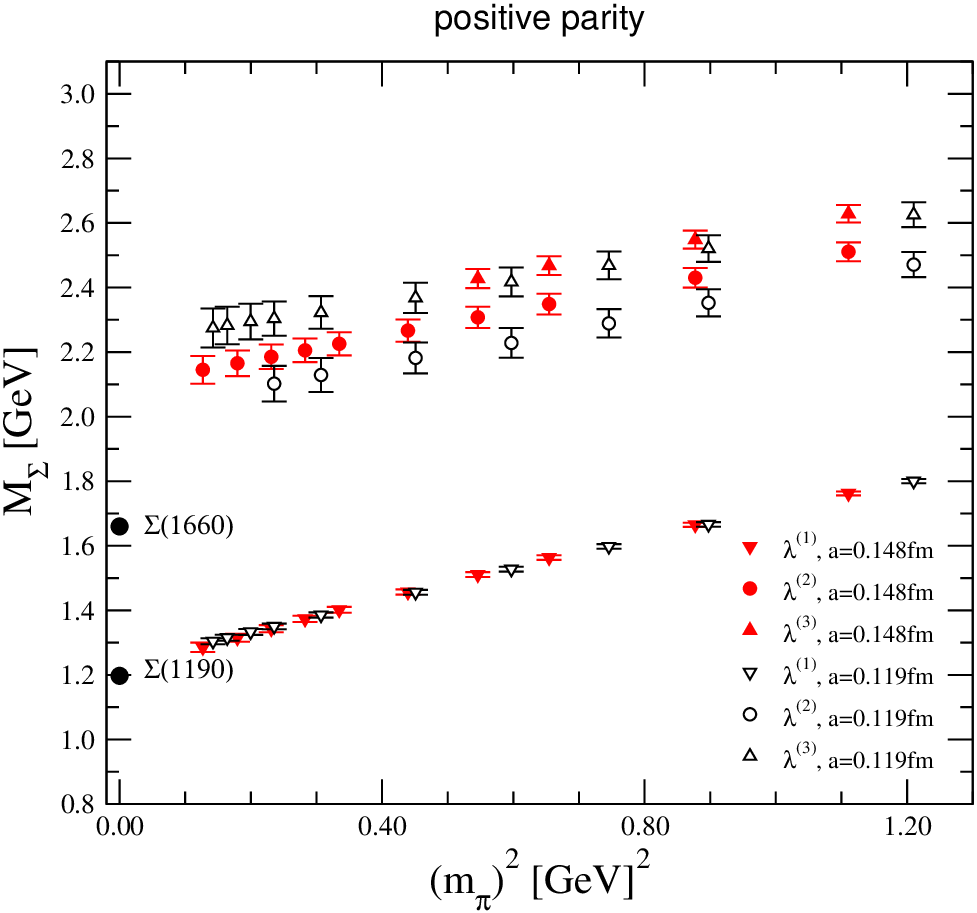}
\hspace*{10mm}
\includegraphics*[height=70mm,clip]{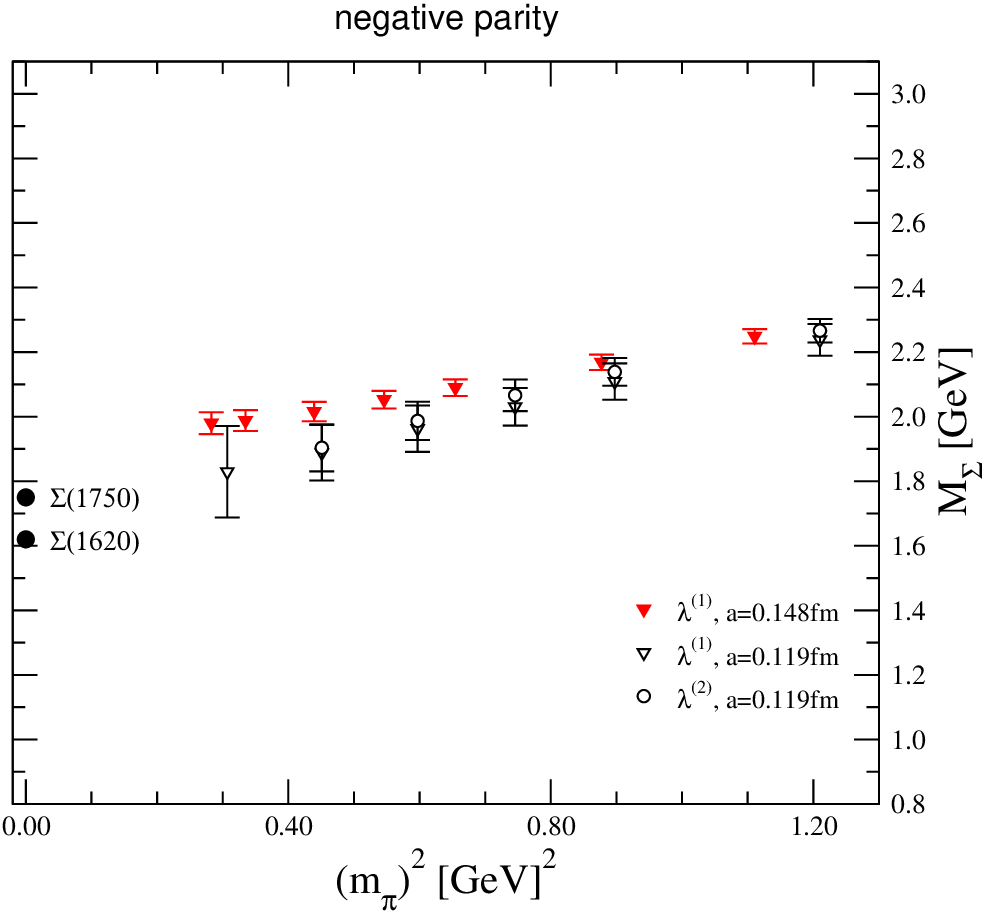}

\vskip5mm

\includegraphics*[height=70mm,clip]{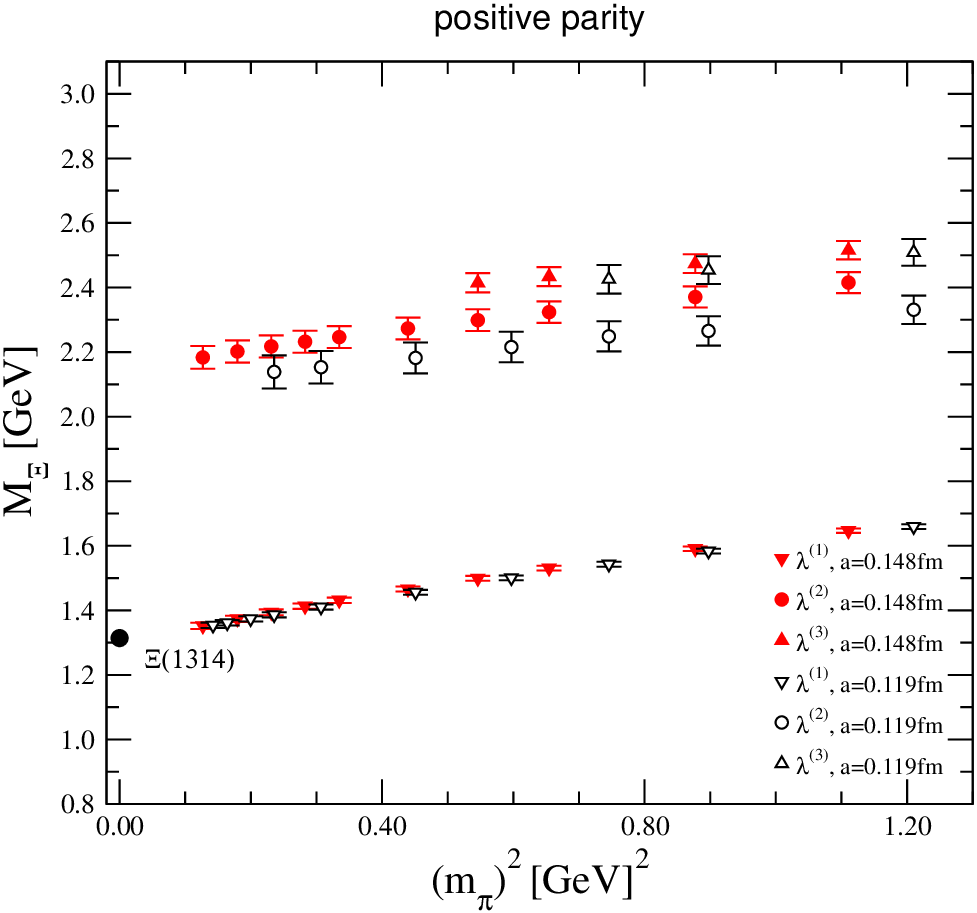}
\hspace*{10mm}
\includegraphics*[height=70mm,clip]{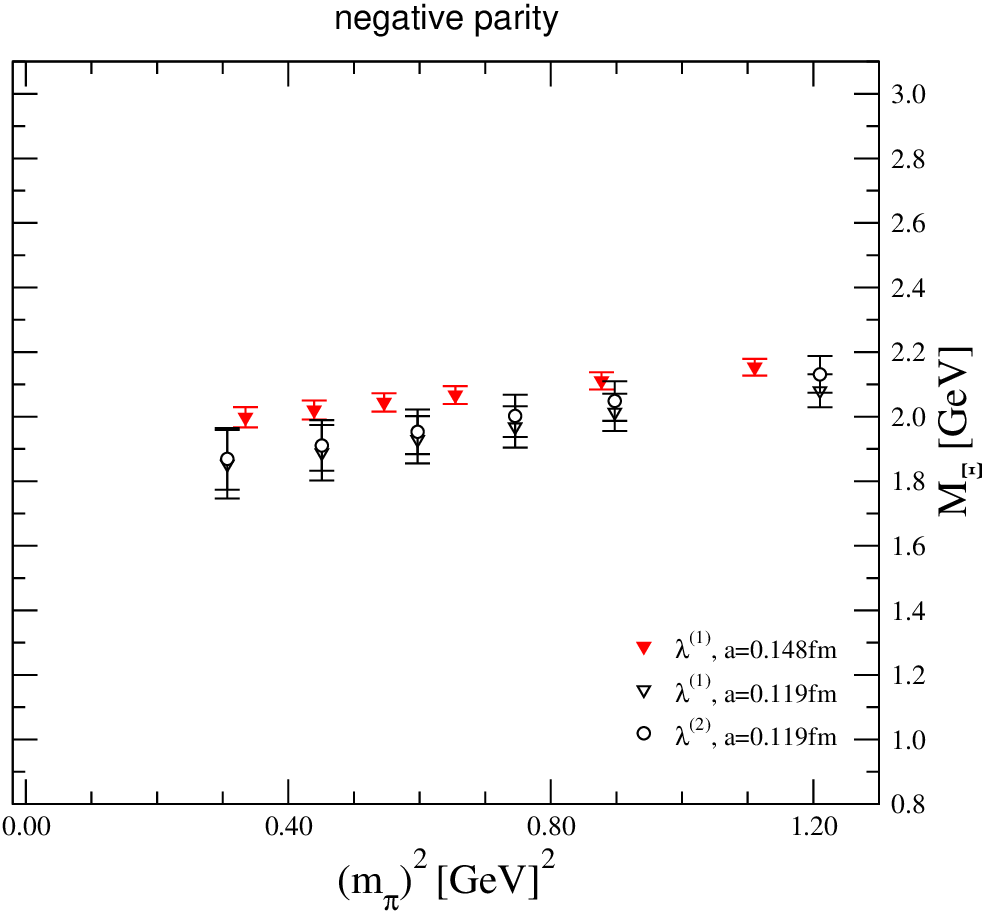}
\end{center}
\caption{Same as Fig.\ \ref{nucleonresults}, now for $\Sigma$ and $\Xi$.} 
\label{sigmaxiresults}
\end{figure*}

Those $\Sigma$ and $\Xi$ resonances which belong to the octet
are structurally identical to the nucleon:
only one and two, respectively, of the light quarks are replaced by a strange
quark. Consequently, their analysis and also the results are only a
variation of what has been found for the nucleon system. We use the same
combination of interpolators in the $6 \times 6$ (for positive parity) and 
$4 \times 4$ (negative parity) correlation matrices as we did for the 
nucleons. 

We present our results for the octet $\Sigma$ and $\Xi$ masses in Fig.\
\ref{sigmaxiresults}. As for the nucleon system, the positive parity $\Sigma$
and $\Xi$ ground states are compatible with the experimental numbers
and essentially no cutoff effects are visible. Concerning the excited positive
parity states, only the first excited states show notable cutoff effects,
while the masses of the second excited states from the two lattices are
compatible within error bars. For the $\Sigma$, where at least the first
excited state is classified, our data extrapolate to a number which is about
20 \% larger than the experimental result, similar to the nucleon case. 

For negative parity, we find two nearly degenerate states which show clear
cutoff effects for the smaller quark masses. For the $\Sigma$ the data are
compatible with the known states. For the negative parity $\Xi$ our data
extrapolate to two states near 1800 MeV (see also the discussion in Subsection F).   

\subsection{$\Lambda$}

\begin{figure*}
\begin{center}
\includegraphics*[height=70mm,clip]{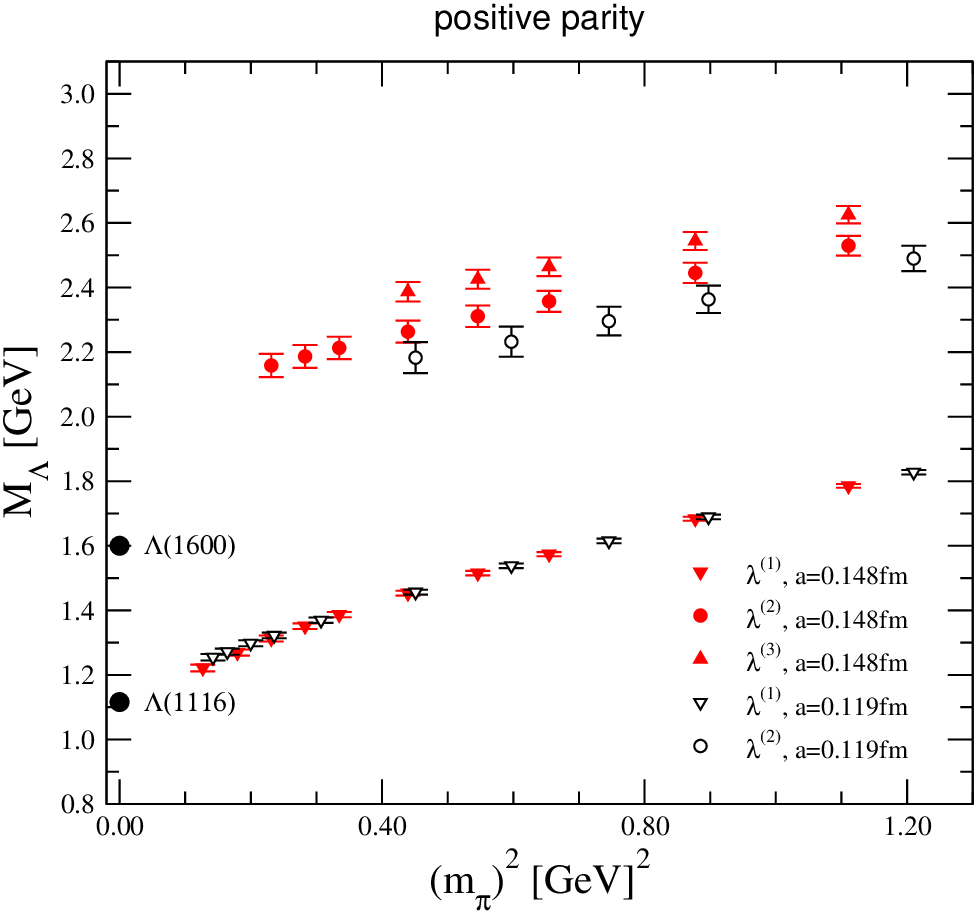}
\hspace*{10mm}
\includegraphics*[height=70mm,clip]{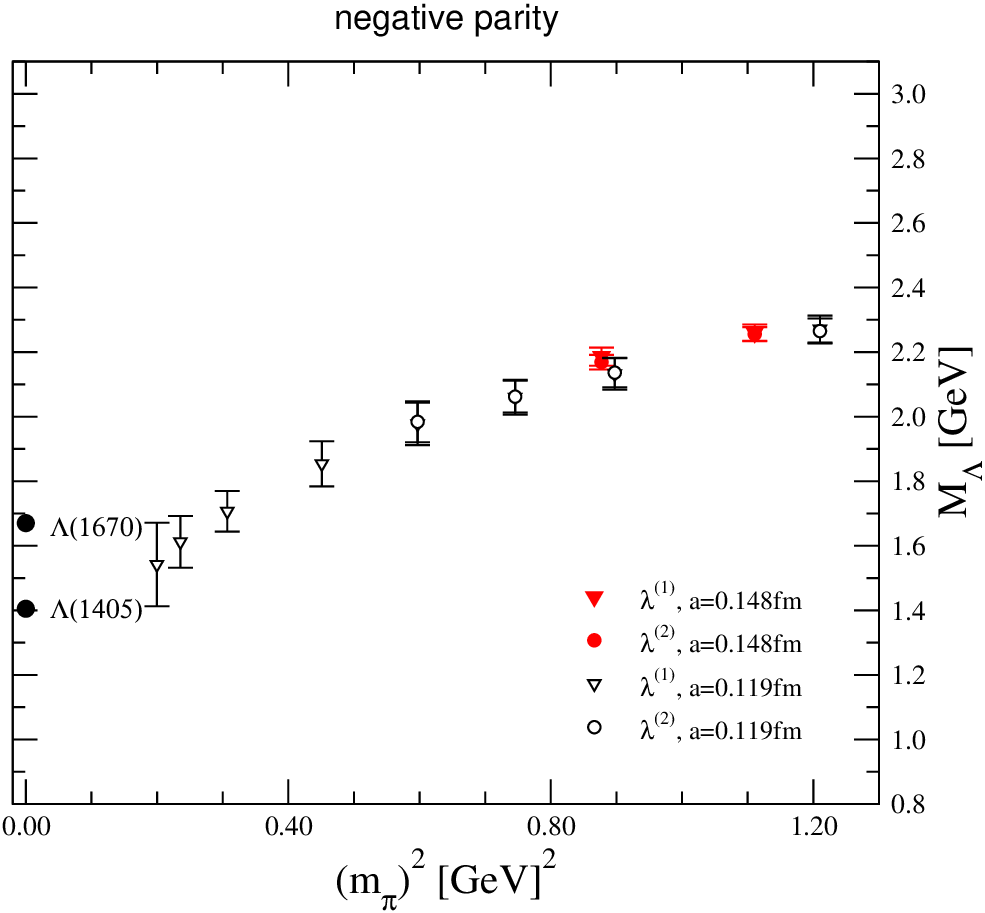}

\vskip5mm

\includegraphics*[height=70mm,clip]{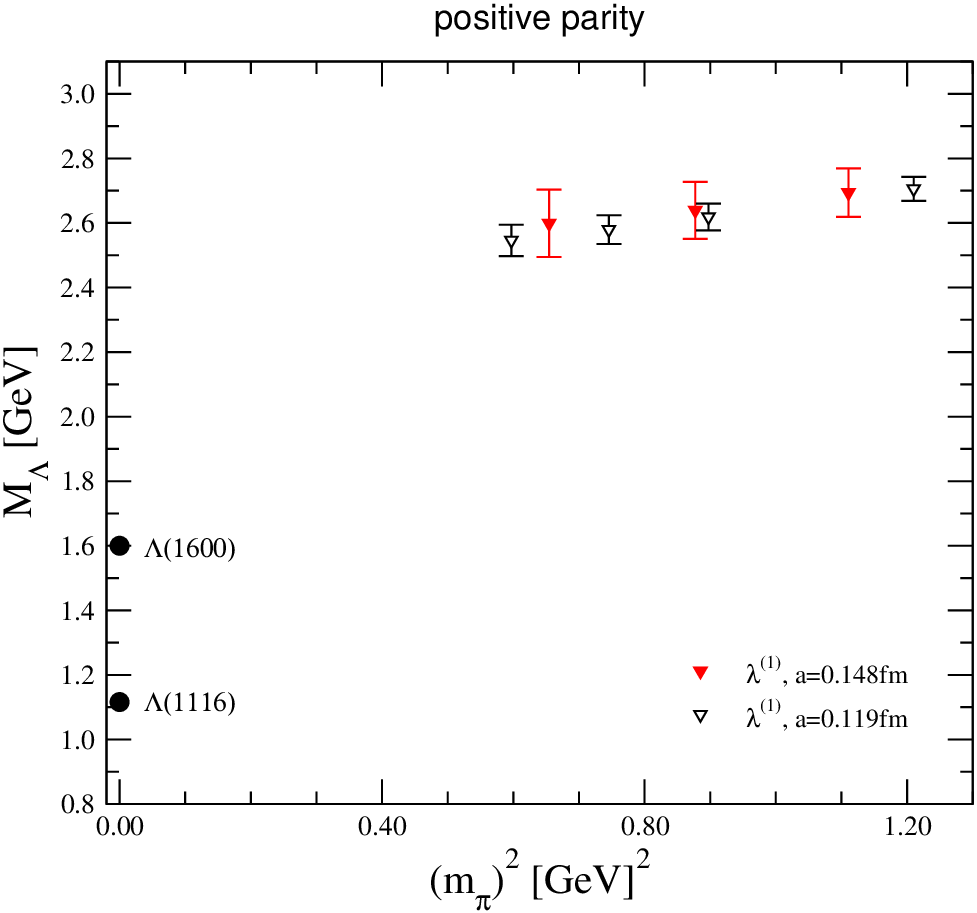}
\hspace*{10mm}
\includegraphics*[height=70mm,clip]{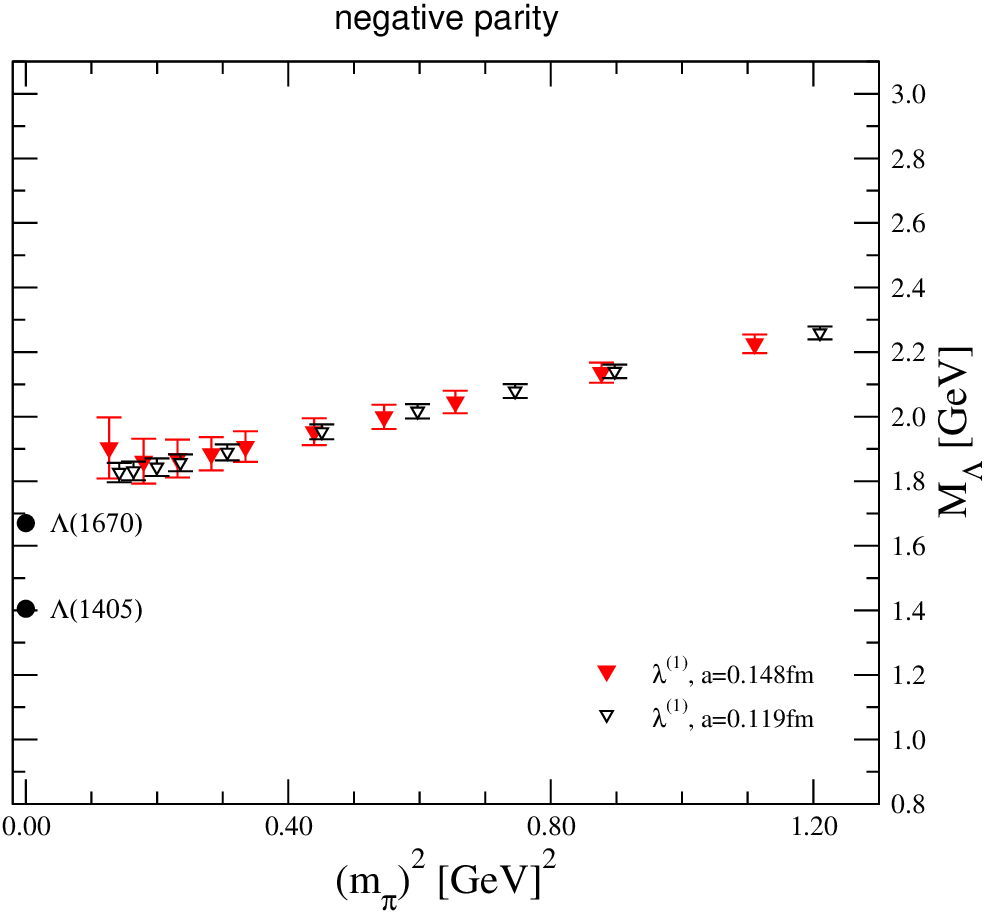}
\end{center}
\caption{
Ground and excited state masses obtained from our $\Lambda$ octet (upper plots) and 
$\Lambda$ singlet (lower plots) interpolators.
}
\label{lambdaresults}
\end{figure*}

For $\Lambda$ we have considered two different kinds of interpolators; one which 
is a pure flavor singlet and one which has mainly overlap with a flavor octet.

For the flavor octet $\Lambda$ we obtain results which are similar to the results 
of the other flavor octet baryons. Even the same combination of sources used
for N, $\Sigma$ and $\Xi$ turns out to be the optimal one also for the $\Lambda$ octet
channel. 

For the flavor singlet $\Lambda$ we are mainly interested in the ground state in both
parity channels. We have therefore used only a single interpolator, the one where all
quarks are smeared narrowly (choosing a different smearing combination does not change 
the results). 

The interesting observation is that while the negative parity flavor-singlet
state extrapolates to the mass which is essentially higher than $\Lambda(1405)$,
which is consistent with previous quenched lattice results, the flavor-octet
negative parity ground state signal is consistent with the $\Lambda(1405)$. Within
the simple quark model picture the negative parity pair $1/2^-, 
\Lambda(1405) - 3/2^-, \Lambda(1520)$
is a flavor-singlet. However, starting from the early Dalitz' work it is also
understood that at least a significant part of $\Lambda(1405)$ could be due to 
$\overline{K} N$
physics \cite{D}. The $\overline{K} N$ bound state system can couple to the flavor-octet
interpolator and our results hint at the  $\overline{K} N$ nature of $\Lambda(1405)$.
It would be very interesting to study the $3/2^-, \Lambda(1520)$ resonance and to see
whether it is a flavor-singlet or flavor-octet state.

\subsection{$\Delta^\frac32$ and $\Omega^\frac32$}

\begin{figure*}
\begin{center}
\includegraphics*[height=70mm,clip]{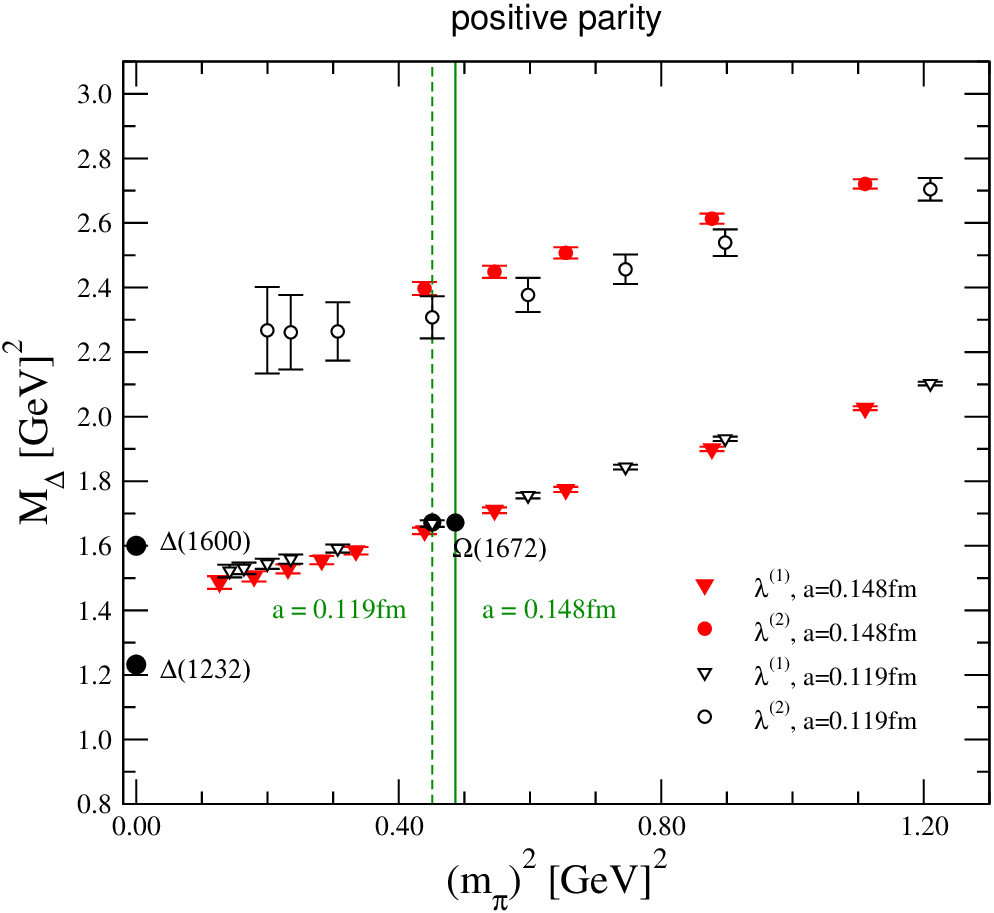}
\hspace*{10mm}
\includegraphics*[height=70mm,clip]{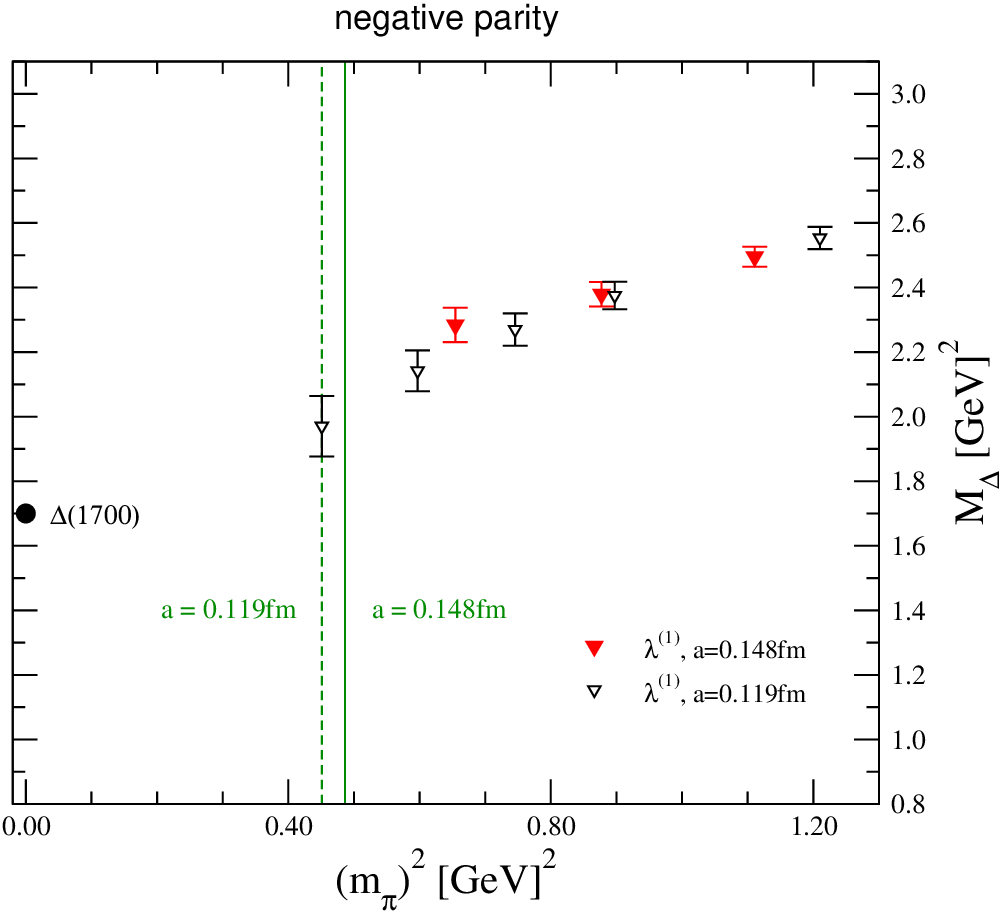}
\end{center}
\caption{
$\Delta$ masses versus $M_\pi^2$.
The vertical lines mark the values of $M_\pi^2$ corresponding to the physical strange
quark mass. 
}
\label{deltaresults}
\end{figure*}

As already discussed in the previous section, our interpolators for $\Delta^\frac32$ 
have to be spin projected to obtain correlators of states with definite quantum numbers.
After the projection we are left with a set of 8 interpolators which differ only in the
smearing combination of the quarks. From these we have chosen different subsets and 
found that the dependence on the chosen subset is only marginal. In the end, we decided
to use the combinations $n(nn)$, $w(nn)$, $n(nw)$, $w(nw)$, 
$n(ww)$, $w(ww)$ for both parity channels. 

In Fig.\ \ref{deltaresults}, we present the results for the $\Delta^\frac32$ and
$\Omega^\frac32$ masses. The positive parity states are shown in the left plot, 
the right plot is for negative parity. The vertical lines in both plots mark 
the values of $m_\pi^2$ corresponding to the physical strange quark mass, which has been 
determined from a fit to the K-meson mass
in a separate calculation on the two lattices. At these values 
of the pion mass we extract the masses for the $\Omega^\frac32$ resonance from our results
for the $\Delta^\frac32$. It is remarkable that the ground state
$\Omega^\frac32$ lies right on top of the experimental value.

The results for the positive parity ground states of $\Delta^\frac32$ show significant 
discrepancies with the experimental results. However, 
this is not unexpected and has already
been observed by other groups \cite{excitedspectro3}. 
The Roper-like state, $\Delta(1600)$,
is not reproduced either. In both cases the most probable explanation would be
a lack of the proper chiral dynamics in quenched QCD. Given the fact that
the $\Omega$ ground state is perfectly reproduced, one concludes that this missing chiral
dynamics becomes especially important at the quark masses below the strange quark mass.

On the negative parity side we could only reliably fit the ground state and only on the
fine lattice do our data reach the strange quark mass such that the mass of the
negative parity $\Omega^\frac32$ can be determined. Extrapolation to the physical limit
is consistent with $\Delta(1700)$.

\subsection{Chiral extrapolations for the fine lattice}

Where the data are sufficient, we perform a chiral extrapolation of our
results. For excited states the form of the chiral extrapolation is not
known from chiral perturbation theory and we extrapolate linearly in
$m_\pi^2$. Since in this paper the focus is on the excited states, 
the extrapolation for the ground states is also kept simple -- we use 
second order polynomials in $m_\pi$ there, which is the structure of the
leading terms in quenched chiral perturbation theory \cite{quenchchpt}. 
Since for some of the states we still observe cutoff effects, we 
extrapolated only the data from the finer lattice.

\begin{figure*}
\begin{center}
\includegraphics*[width=0.47\textwidth,clip]{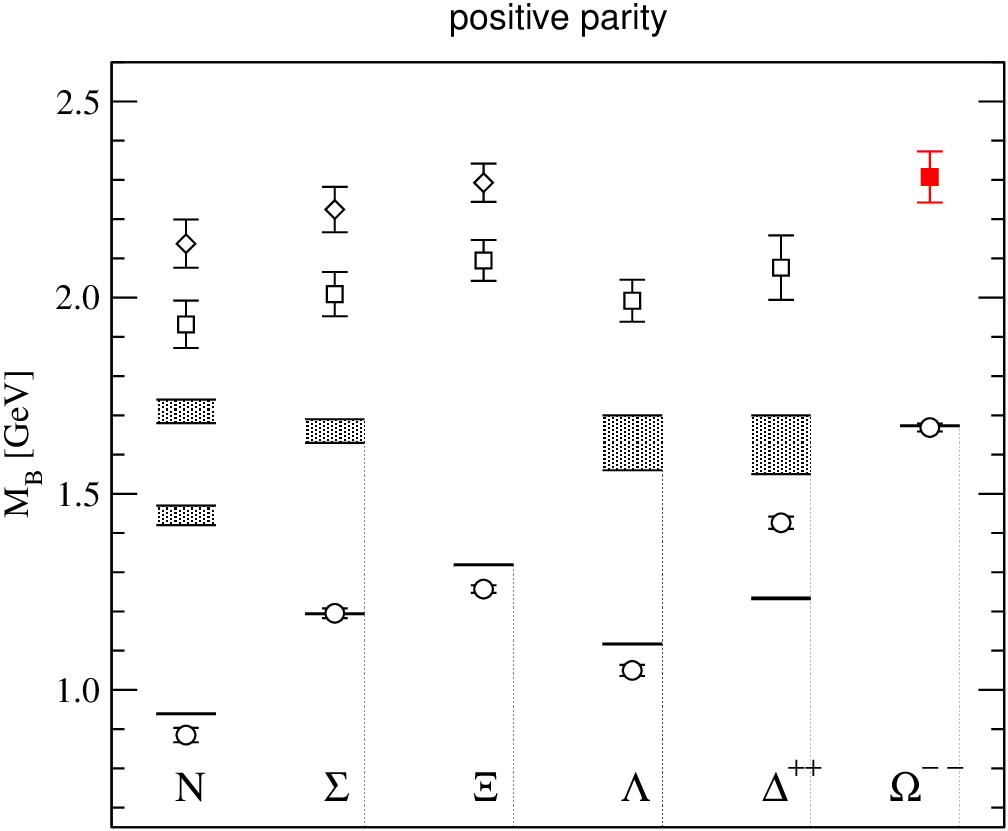}
\hspace{5mm}
\includegraphics*[width=0.47\textwidth,clip]{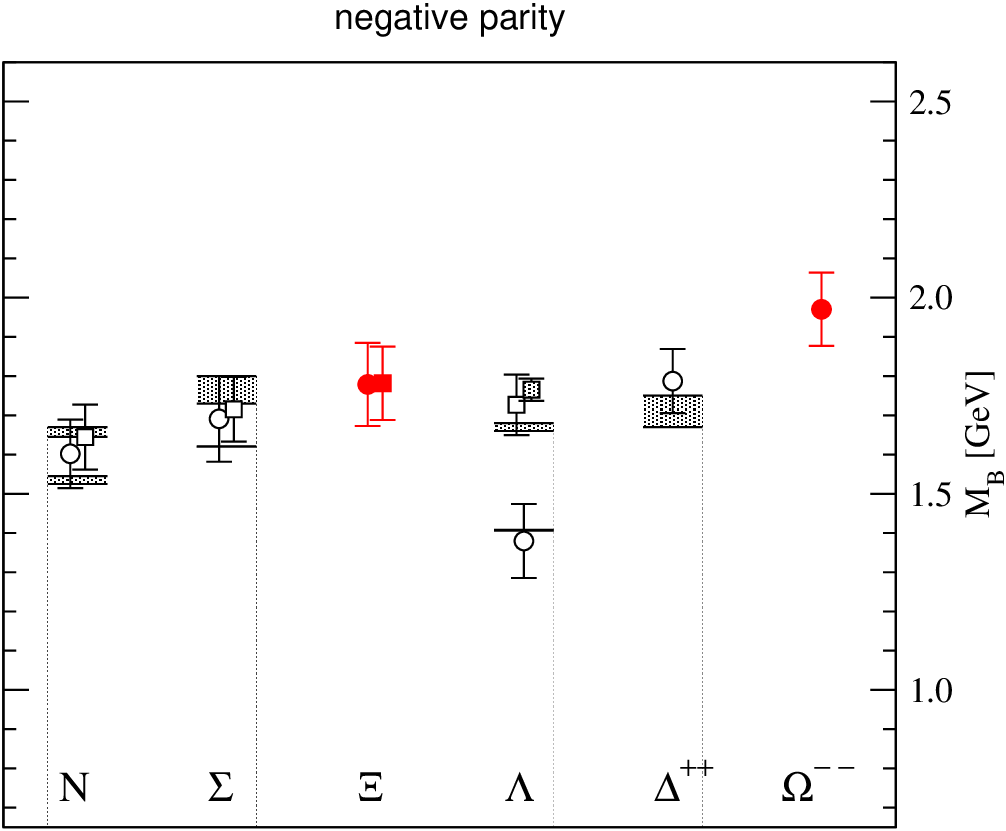}
\end{center}
\caption{
Chiral extrapolation of our results. The left plot is for positive
parity, the right for negative parity. The horizontal bars represent
the experimental numbers (where known), indicating also the error. For our
results we use circles for ground states, squares and diamonds for the first 
and second excited states respectively. The shaded square symbol for the
excited $\Lambda$ represents the chiral limit of the data from the singlet
interpolator. Filled symbols are used for those states where no corresponding
state is listed in the particle data book.
}
\label{chiralextra}
\end{figure*}

For positive parity the results of the chiral extrapolation 
are presented in the left plot of Fig.~\ref{chiralextra}. 
We remark, that the numbers for the $\Omega$ are obtained by an
interpolation to the strange quark mass. While the ground states come 
out reasonably well for a quenched calculation, the results for the 
excited states are systematically 20 \% - 25\% above the experimental numbers (where
known). The most likely explanation is that quenching removes some
important piece of chiral physics, which is actually responsible
for the proper mass of excited positive parity states. Significant
finite volume effects cannot be ruled out either.

For negative parity states (right plot of Fig.~\ref{chiralextra}) 
the results are compatible with the experimental
numbers (where known), although the statistical errors are larger. Also here
we cannot exclude that cutoff effects push our numbers up a little bit, but from
the comparison of the results on our two lattices we estimate that this effect
is not larger than the statistical error. Again the result for $\Omega$ is
obtained from an interpolation to the strange quark mass. 
One may expect that quenching effects are essentially smaller for the negative parity
channel states than for the positive parity excited states. This is expected
a-priori, since all low-lying negative parity states 
have the mixed flavor-spin symmetry {\it[21]}$_{FS}$ and hence are affected by the
chiral dynamics only slightly (except for the $\Lambda(1405)$ ) \cite{GR}.

\section{Summary}
In this article we presented a quenched spectroscopy calculation of excited baryons
using the variational method. We use interpolators with different Dirac
structures. Furthermore each quark can either have a narrow or a wide source
such that the states can have nodes in their spatial wave function.

For the positive parity baryons we find that the ground state masses are compatible
with the experimental numbers, while for the excited states the masses are systematically
20\% - 25\% above the experimental numbers.  We believe that the
failure to reproduce the masses of the positive parity excited baryons is
indeed mainly due to quenching where a significant part of chiral physics
is missing. Large finite volume effects cannot be excluded either. 

For negative parity, we find that our masses are in reasonable agreement with
the experimental numbers, although here our statistical errors are larger and
a further lowering of our results for a lattice with a higher cutoff cannot be
excluded.

In some of the channels we analyze, the corresponding baryons are not yet
classified \cite{pdg}. 
For four of these channels we believe that our data are strong
enough to quote the final results as a prediction: The first excited positive parity
$\Omega$ state, the negative parity $\Omega$ ground state, and the ground and
first excited negative parity $\Xi$ states. 

The two $\Omega$ states are
included in this list since at the strange quark mass the chiral dynamics is
less important and also our results do not need to be extrapolated to the
chiral limit. Concerning the two negative parity $\Xi$ states we believe that
the good results of the structurally very similar negative parity nucleons and 
$\Sigma$ baryons justify the prediction of the mass of the negative parity
ground and first excited state in the $\Xi$ channel. Our final numbers for the
masses of the four states are listed in Table~\ref{predictions}.

\begin{table}[t]
\begin{center}
\begin{tabular}{l|c}
\hspace{20mm} state \hspace{7mm} & \hspace{7mm} Mass [MeV] \hspace{7mm} \\
\hline 
$\Omega$, positive parity, first excited state & 2300(70) \\
$\Omega$, negative parity, ground state & 1970(90) \\
$\Xi$, negative parity, ground state & 1780(90) \\
$\Xi$, negative parity, first excited state & 1780(110) \\
\end{tabular}
\end{center}
\caption{Collection of our final results for some states not classified by the
  Particle Data Group \cite{pdg}.}
\label{predictions}
\end{table}

\begin{acknowledgments}
C.G.\ acknowledges interesting discussions on the variational method 
with Rainer Sommer and Peter Weisz.
The calculations were performed on the Hitachi SR8000 
at the Leibniz Rechenzentrum in Munich and the JUMP Cluster at the 
Zentralinstitut f\"ur angewandte Mathematik in J\"ulich. We thank the 
LRZ and ZAM staff for training and support. L.Y.G.\ is supported by 
``Fonds zur F\"orderung der Wissenschaftlichen Forschung in \"Osterreich'', 
FWF, project P16823-N08. This work is supported by DFG and BMBF. C.G.\ and
C.B.L.\ thank the Institute for Nuclear Theory at the University of 
Washington for its hospitality and the Department of Energy for partial 
support during the completion of this work. 
\end{acknowledgments}

\end{document}